\documentstyle[twocolumn,epsfig,prl,aps,amsmath,amssymb]{revtex}
\begin{document}
\draft
\twocolumn[\hsize\textwidth\columnwidth\hsize\csname
@twocolumnfalse\endcsname
\title{Cluster Monte Carlo Algorithm for the Quantum Rotor Model}
\author{Fabien Alet$^{a,\dagger}$ and Erik~S.~S\o rensen$^b$}
\address{
$^a$Laboratoire de Physique Quantique \& 
UMR 5626, Universit\'e Paul Sabatier, 31062 Toulouse, France\\
$^b$Department of Physics and Astronomy, McMaster University, Hamilton,
ON, L8S 4M1 Canada}  
\date{\today}
\maketitle
\begin{abstract}
We propose a highly efficient ``worm" like cluster Monte Carlo algorithm
for the quantum rotor model in the link-current representation.
We explicitly prove detailed balance for the new algorithm even in the presence of disorder. For 
the pure quantum rotor model with $\mu=0$ the new algorithm yields high precision
estimates for the critical point $K_c=0.33305(5)$ and the correlation length
exponent $\nu=0.670(3)$. For the disordered case, $\mu=\frac{1}{2}\pm\frac{1}{2}$, we find $\nu=1.15(10)$.
\end{abstract}
\pacs{ 74.20.Mn, 73.43.Nq,  02.70.-c }
\vskip2pc]

What types of insulating, conducting, superconducting and more exotic phases
occur in two-dimensional systems at $T=0$ is a topic of considerable current
interest. A significant amount of
theoretical~\cite{fisher89c,fisher90a,damle,bmetal} and
experimental~\cite{exp,mason} effort has focused on bosonic systems where a
superconductor to insulator transition is known to occur.  In agreement with
most experiments~\cite{exp}, it was under quite general conditions
shown~\cite{fisher89c,fisher90a} that a transition can occur directly between
the superconducting and insulating states.  However, more recently, it has been
suggested that an exotic {\it metallic} phase also is
possible~\cite{bmetal,mason}.  In this context precise numerical results would
be very valuable and in the present paper we propose a new, very efficient
cluster Monte Carlo algorithm for this purpose, allowing us to significantly
improve previous results. In particular we show that the
inequality~\cite{Chayes86} $\nu\ge 2/d$ is {\it not} violated in the presence of disorder, resolving contradictions in previous work.
The high precision of the algorithm should allow for precise calculations
of transport properties of quantum rotor models studied theoretically
in~\cite{fisher89c,bmetal}. The ideas presented here could be useful for the
study of classical spin systems~\cite{muquarter}.

Low-dimensional bosonic systems are often described in terms
of the (disordered) boson Hubbard model: 
$
H_{\rm bH}=\sum_{\bf r}\left(\frac{U}{2}\hat n_{\bf r}^2 -
\mu_{\bf r}\hat n_{\bf r}\right)- t_0\sum_{\langle {\bf r},{\bf r'}\rangle }
(\hat \Phi^\dagger_{\bf r}\hat \Phi_{\bf r'}+c.c) \ .
$
Here $U$ is the on-site repulsion, $t_0$ the hopping strength, 
$\mu_{\bf r}$ the chemical potential varying
uniformly in space between $\mu\pm\Delta$ and
$\hat n_{\bf r}=\hat \Phi^\dagger_{\bf r}\hat \Phi_{\bf r}$ is the number operator.
If we set $\hat \Phi_{\bf r}\equiv|\hat \Phi_{\bf r}|e^{i\hat \theta_{\bf r}}$ and
integrate out amplitude fluctuations, $H_{\rm bH}$ becomes equivalent 
to the quantum rotor model~\cite{long}:
\begin{equation}
H_{\text{qr}}=
\frac{U}{2}\sum_{\bf r}
\left( \frac{1}{i}\frac{\partial}{\partial
\theta_{\bf r}} \right)^2 
+i\sum_{\bf r} \mu_{\bf r}
\frac{\partial}{\partial \theta_{\bf r}}
-t\sum_{\langle {\bf r},{\bf r'}\rangle }
\cos(\theta_{\bf r}-\theta_{\bf r'}).
\label{eq:hqr}
\end{equation}
Here, 
$\theta_{\bf r}$ the phase of the quantum rotor, $t$ the renormalized
hopping strength and
$\frac{1}{i}\frac{\partial}{\partial\theta_{\bf r}}
\simeq n_{\bf r}$. 
The quantum rotor model describes a wide range of phase transitions dominated by phase-fluctuations
and it is well known~\cite{long} that an equivalent classical
model exists where the Hamiltonian is written in terms of currents defined on the links of
a lattice, ${\bf J}=(J^x,J^y,J^\tau)$. These link-current variables describe the 
``relativistic" bosonic current which should be divergenceless, ${\bf \nabla \cdot J} = 0$. 
In terms of these variables the classical (2+1)D Hamiltonian can be
written as follows~\cite{long}:
\begin{equation}
H=\frac{1}{K} {\sum_{({\bf r},\tau)}}^\prime
\left[\frac{1}{2} {\bf J}_{({\bf r},\tau)}^{2}- \mu_{\bf r} J_{({\bf r},\tau)}^\tau\right].
\label{eq:hV}
\end{equation}
$\sum^\prime$ denotes a summation over configurations with ${\bf \nabla \cdot J} = 0$.
Varying $K$ corresponds to changing the ratio $t/U$ in the quantum rotor
model. The quantum rotor model has been extensively
studied~\cite{long,cha,others} in this representation, but a number of
conclusions can be questioned due to severe finite-size effects.
For notational convenience it is useful to slightly enlarge 
the definition of the link-currents at a given site in the following way:
At each site $({\bf r},\tau)$ in the lattice we define {\it six} surrounding
link variables $J_{({\bf r},\tau)}^\sigma$ where $\sigma$ runs over $\pm x,\pm
y,\pm \tau$.
Note that, with this notation $J^{-x}_{(x,y,\tau)}$ and $J^x_{(x-1,y,\tau)}$ is the same
variable, with
equivalent relations in the other directions.
The divergenceless constraint at the site $({\bf r},\tau)$ 
can then be written:
$J^{-x}+J^{-y}+J^{-\tau}=J^x+J^y+J^\tau$,
so that the sum of the incoming
and outgoing currents are equal.
Conventional Monte Carlo updates~\cite{long} on this model consists of updating
simultaneously four link variables.
Global moves, updating a whole line of link variables thus allowing particle
and winding numbers to fluctuate, are added
to ensure ergodicity, but the acceptance ratio for these
moves becomes exponentially small for large lattice sizes.
Here we will describe a way to construct a
worm-like algorithm to perform non local moves for this model.

The cluster algorithm \cite{Swendsen,wolff,loop} we propose is similar in
spirit to worm algorithms \cite{Prokofev,Sandvik} in the sense that we
update the link-currents by moving a ``worm'' through the lattice visiting the 
sites $s_i=({\bf r}_i,\tau_i)$. 
The links through which the worm pass
are updated {\it during} its construction. 
At a given site, the links with $\sigma$ equal to $x,y,\tau$ are called 
outgoing links and those with $\sigma$  equal to $-x,-y,-\tau$ incoming links.
When the worm is moving through the lattice the currents $J^\sigma_{s_i}$ 
are updated in the following manner:
if the worm is leaving the site $s_i$ along an outgoing link we {\it increment}
the corresponding current:
\begin{equation}
J^\sigma_{s_i}\rightarrow J'^\sigma_{s_i} = J^\sigma_{s_i}+1,\ \
\sigma=x,y,\tau.
\label{eq:rout}
\end{equation}
If the worm is leaving the site $s_i$ along an incoming link we {\it decrement}
the corresponding current:
\begin{equation}
J^\sigma_{s_i} \rightarrow J'^\sigma_{s_i} = J^\sigma_{s_i}-1,\ \
\sigma=-x,-y,-\tau.
\label{eq:rin}
\end{equation}
The construction of the worm starts with the choice of a random initial site
$s_1=({\bf r}_1,\tau_1)$ in the space-time lattice. Then the algorithm can
be decomposed in two steps. $(i)$ The
worm moves to one of the 6 neighboring sites. To decide
which direction to go from a site $s_i=({\bf r}_i,\tau_i)$, we calculate for all directions
$\sigma=\pm x,\pm y,\pm \tau$ weights, $A^\sigma_{s_i}$, according to local detailed balance. A good choice is: 
\begin{equation}
A^\sigma_{s_i}
=\min(1,\exp(-\Delta E_{s_i}^{\sigma}/K)),\ \ 
\Delta E_{s_i}^{\sigma}=E'^\sigma_{s_i}-E_{s_i}^{\sigma}.
\end{equation}
Here $E^\sigma_{s_i}=\frac{1}{2}(J^\sigma_{s_i})^2-\mu_{\bf r_i}
J^\sigma_{s_i}\delta_{\sigma,\pm \tau}$ is the contribution to the total
energy from the link $J^\sigma_{s_i}$, before the worm moves through it. $E'^\sigma_{s_i}$ is the energy contribution with
$J^\sigma_{s_i}$
replaced by $J'^\sigma_{s_i}$. 
By normalizing the $A^\sigma_{s_i}$'s we define the probabilities:
$
p^\sigma_{s_i}={A^\sigma_{s_i}}/{N_{s_i}},
$
where $N_{s_i}=\sum_\sigma A^\sigma_{s_i}$.  A direction $\sigma$ is
then chosen according to these probabilities. $(ii)$ Once $\sigma$ is chosen, we update the
corresponding $J^\sigma_{s_i}$ according to the above rules,
Eq.~(\ref{eq:rout}) and (\ref{eq:rin}), and extend the worm to
the new lattice site $s_{i+1}$. $(i)$ and $(ii)$ are then repeated until 
the worm passes through the initial site where $s_{i+1}=s_1$.
Finally, in order to satisfy detailed balance we have to {\it erase} the
worm with a probability determined in the following way.
If $N$(worm) and $N$(no worm) are the normalization of the probabilities at the initial site $s_1$ {\it with}
and {\it without} the worm present, then we erase the constructed
worm with a probability 
\begin{equation}
P^e=1-{\rm min}(1,\frac{N(\rm no\ worm)}{N(\rm worm)}).
\label{eq:perase}
\end{equation}
Under most conditions this probability is very small.
Several points are noteworthy about this algorithm. First of all, the
configurations generated during the construction of the worm are not valid
(${\bf \nabla\cdot J}\neq 0$). However, once the construction of the worm is
finished and the path of the worm closed, the divergenceless constraint is
satisfied. Secondly, when the worm moves through the lattice it may pass
many times through the same link and cross itself before it reaches the
initial site where the construction terminates. Hence, it is crucial that
the current variables are updated {\it during} the construction of the worm.
Finally, at each step $i$ in the construction of the worm it is
likely that the worm at the site $s_i$ will partially ``erase" itself by
choosing to go back to the site $s_{i-1}$ visited immediately before, 
thereby ``bouncing" off the site $s_i$.

Now we turn to the proof of detailed balance for the
algorithm.  Let us consider the case where the
worm, $w$, visits the sites $\{s_1\ldots s_N\}$ where $s_1$ is the initial
site. The worm then goes through the corresponding link variables
$\{l_1\ldots l_N\}$, with $l_i$ connecting $s_{i}$ and $s_{i+1}$.
Note that $s_N$ is the last site visited before the worm reaches
$s_1$. Hence, $s_N$ and $s_1$ are connected by the link $l_N$.
The total probability for constructing the worm $w$ is then given by:
$
P_w = P_{s_1}(1-P_{w}^e)\prod_{i=1}^N A_{s_i}^{\sigma}/{N_{s_i}}.
$
The index $\sigma$ denotes the direction needed to go from
$s_i$ to $s_{i+1}$,
$P_{s_1}$ is the probability for choosing site $s_1$ as the starting
point and $P_{w}^e$ is the probability for erasing the worm after
construction. If the worm $w$ has been accepted we have to consider the
probability for reversing the move. That is, we consider the probability for constructing an anti-worm
$\bar w$ annihilating the worm $w$. We have:
$
P_{\bar w} = 
P_{\bar s_1}(1-P_{\bar w}^e)
\prod_{i=1}^N{{\bar A}_{\bar s_i}^{\sigma}}/{{\bar N}_{\bar s_i}}.
$
Here, the index $\sigma$ denotes the direction needed to go from
$\bar s_i$ to $\bar s_{i+1}$.
Note that, in this case the sites are visited in the opposite order,
$\bar s_1 = s_1, \bar s_2 = s_{N},
\ldots, \bar s_N = s_2$, in general $\bar s_i=s_{N-i+2}\ (i\neq 1)$. 
Note also that, $\bar s_i$ and $\bar s_{i+1}$ are connected by the link
$\bar l_i=l_{N-i+1}$
with $\bar s_N$ and $\bar s_1$ connected by $\bar l_N=l_1$. 
With this notation we see that,
$s_i$ and $\bar s_{N-i+1}\equiv s_{i+1}$ are connected by the link variable
$l_i$.
Let us now consider the case where both of the worms $w$ and $\bar w$ have reached the site
$s_i$ {\it different} from the starting site $s_1$. Since we are updating the link variables during
the construction of the worm and since we are always considering moving
the worm in all six
directions, we have $N_{s_i}=\bar N_{\bar s_{N-i+2}= s_i}\ \ (i\neq 1)$.
Furthermore, $A_{s_i}^{\sigma}$ and
$\bar A_{\bar s_{N-i+1}= s_{i+1}}^{\sigma}$ only
depend on the link variable $l_i$ connecting the sites $s_i$ and $\bar
s_{N-i+1}$ and 
we see that:
$
{A_{s_i}^{\sigma}}/
{\bar A_{\bar s_{N-i+1}}^{\sigma}}=
\exp(-\Delta E_{s_i}^{\sigma}/K),\ i=1\ldots N.
$
Hence, since $P_{s_1}=P_{\bar s_1}$, we find:
\begin{equation}
\frac{P_w}
{P_{\bar w}}
=\frac{1-P_{w}^e}{1-P_{\bar w}^e}
\frac{\bar N_{\bar s_1}}{N_{s_1}}
\exp(-\Delta E_{\rm Tot}/K),
\end{equation}
where $\Delta E_{\rm Tot}$ is the total energy difference between a
configuration with and without the worm $w$ present. Now we consider 
$P^e=1-\min(1,{N_{s_1}(\rm no\ worm)}/{N_{s_1}(\rm worm)})$.
Here, $N_{s_1}$$\equiv$$N_{s_1}$(no worm) is equal to  $\bar
N_{s_1}$(anti-worm) and $\bar N_{s_1}$(no anti-worm)$\equiv$$\bar
N_{s_1}$ is equal to $N_{s_1}$(worm).
Hence, we find for the probability to erase the worm
$P_{w}^e=1-\min(1,N_{s_1}/\bar N_{s_1})$ and $P_{\bar w}^e=1-\min(1,\bar N_{s_1}/N_{s_1})$ for erasing the anti-worm.
With this choice of $P^e$ we satisfy detailed balance since:
$
\frac{P_w}{P_{\bar w}}=
\exp(-\Delta E_{\rm Tot}/K).
$
Ergodicity is simply proven as the worm can
perform local loops  and wind around the
lattice in any direction, as in the conventional algorithm.
\begin{figure}
\begin{center}
\epsfig{file=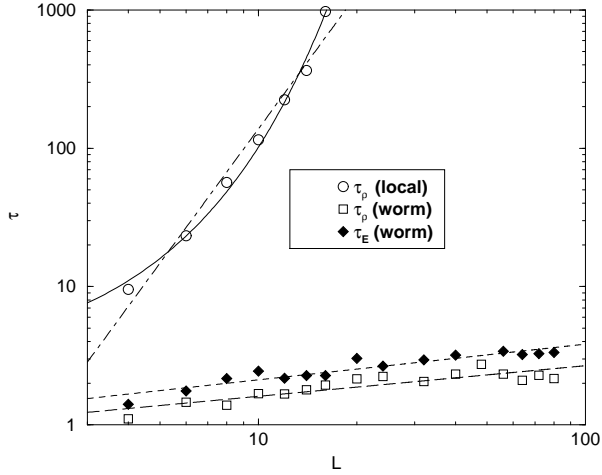,width=8cm}
\caption{Autocorrelation times versus lattice size
for the conventional and worm algorithm for $\mu=0$ at $K=0.333$. The
dashed lines indicate power-law fits and the solid line an exponential
fit in $L$.}
\label{fig:fig1}
\end{center}
\end{figure}

To demonstrate the efficiency of the proposed algorithm we have calculated
auto-correlation times for different lattice sizes for the worm
algorithm and the conventional algorithm. For an observable
${\cal O}$ we define the auto-correlation function and the
auto-correlation time $\tau_{\cal O}$ in the usual
manner~\cite{newman}:
\begin{equation}
\frac{\langle{\cal O}(0){\cal O}(t)\rangle-\langle{\cal O}\rangle^2}{
\langle{\cal O}^2\rangle-\langle{\cal O}\rangle^2} =  
ae^{-t/\tau_1}+
be^{-t/\tau_{\cal O}}+\ldots.
\end{equation}
Here, $t$ is the Monte Carlo time measured in Monte Carlo sweeps
(MCS), with 1 MCS corresponding to $L^d$ attempted updates. 
The auto-correlation function is calculated from simulations with 10$^8$
MCS, and to obtain the best estimate of $\tau_{\cal O}$ we
always fit to the indicated double-exponential form with $\tau_1 \ll
\tau_{\cal O}$.
To make a fair comparison of $\tau_{\cal O}$ for the two
algorithms, one customarily~\cite{wolff,newman} multiplies $\tau_{\cal O}$ for
the worm algorithm  by $N/\left< l \right>$, with $\left< l \right>$ the
mean number of links in a worm and $N=3L^3$. 
With this rescaling we show in Fig.~\ref{fig:fig1} the auto-correlation
times, $\tau_\rho$ for the stiffness (see
exact definition below) at $\mu=0$ for both algorithms. 
The calculations have been performed on cubic lattices at $K=0.333$, 
very near previous estimates of the critical point~\cite{cha}. 
For the worm algorithm we also show the auto-correlation time for the
energy, $\tau_{E}$, which is almost identical to $\tau_\rho$.
The auto-correlation times increase dramatically with system size
for the conventional algorithm where as they remain very small (of the order
of 2-3 MCS per link) for the worm algorithm. If we fit
the $L$ dependence of $\tau_\rho\sim L^{z_{MC}}$ with a power law, we
obtain an auto-correlation
exponent $z_{MC}$ larger than $4$ for the conventional algorithm. For
the conventional algorithm it is likely that $\tau_\rho$ is diverging
exponentially with $L$ since $\rho$ is solely determined by
global updates for which the acceptance probability decreases
exponentially with $L$. For the worm algorithm we find a very
small $z_{MC} \sim 0.3$. 
\begin{figure}
\begin{center}
\epsfig{file=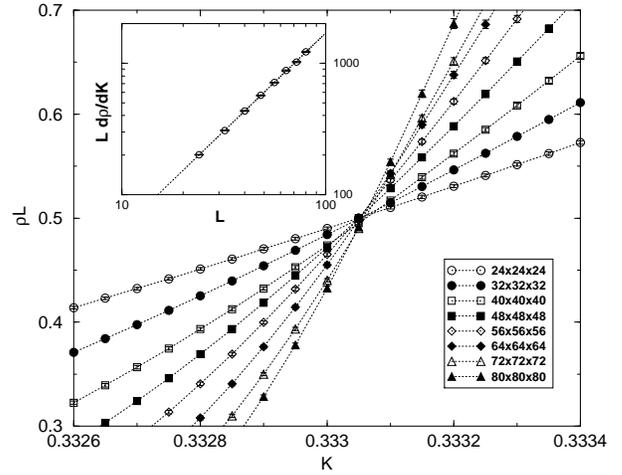,width=8cm}
\caption{$L \rho$ versus $K$ for different lattice sizes, for $\mu=0$. All curves cross
at the critical point $K_c=0.33305(5)$ with $L\rho|_{K=K_c}=0.495(5)$. Inset: $L d\rho/dK$
at $K_c$ versus $L$. The dashed line indicates a fit yielding an exponent
$\nu=0.670(3)$.}
\label{fig:fig2}
\end{center}
\end{figure}

We now present results for
the model Eq.~(\ref{eq:hV}) at $\mu=0$.
There, the model is expected to undergo a transition in the (2+1)D XY
universality  class~\cite{fisher89c,cha} from a
superfluid into a Mott insulating phase with a
dynamical critical exponent $z=1$. The different phases can be distinguished
by calculating the stiffness defined as \cite{long} :
\begin{equation}
\rho = \frac{1}{L_\tau L^2} \langle (\sum_{{\bf r},\tau} J_{({\bf r},\tau)}^x)^2 \rangle.
\end{equation}
Since we expect $z=1$, we use $L_\tau$, the system size in the third
direction, equal to $L$.
To obtain the $K$ dependence of $\rho$ we have
used reweighting techniques~\cite{Ferrenberg88} on large runs
(of the order of 10$^8$ MCS) at $K=0.333$. 
The error bars are determined using jackknife techniques~\cite{newman}.
Using finite-size scaling relations, the quantity $\rho L^z$ is
expected to be independent of system size at the critical point~\cite{long}, $K_c$. 
Moreover, $L^zd\rho/dK$ is expected to diverge at $K_c$ 
as $L^{1/\nu}$ where $\nu$ is the correlation length exponent. We have
explicitly calculated this quantity by evaluating the thermodynamic 
derivative of $\rho$ with respect to the coupling $K$:
$
d\rho/dK = ( \langle \rho E \rangle - \langle \rho
\rangle \langle E \rangle)/K^2.
$
In Fig.~\ref{fig:fig2}, we show $L\rho$ versus K for different lattice
sizes. From the crossing of the curves we can determine $K_c=0.33305(5)$
to a much higher precision than was possible using the conventional
algorithm on much smaller systems~\cite{long,cha,others}. 
Since all the curves cross in a single point our results are completely
consistent with a dynamical exponent $z=1$, as expected~\cite{fisher89c}.
In the inset of Fig.~\ref{fig:fig2}
is shown the size dependence of $L^zd\rho/dK$ at
$K_c$ on a log-log scale. We fit this curve to a power-law $A
L^{1/\nu}$ and obtain $\nu=0.670(3)$, in perfect agreement with
estimates for the three-dimensional XY universality class~\cite{Campostrini01}. 
Preliminary results~\cite{muquarter} for the {\it generic} transition at $\mu=\frac{1}{4}$
show pronounced finite-size effects questioning previous work~\cite{others}.

\begin{figure}
\begin{center}
\epsfig{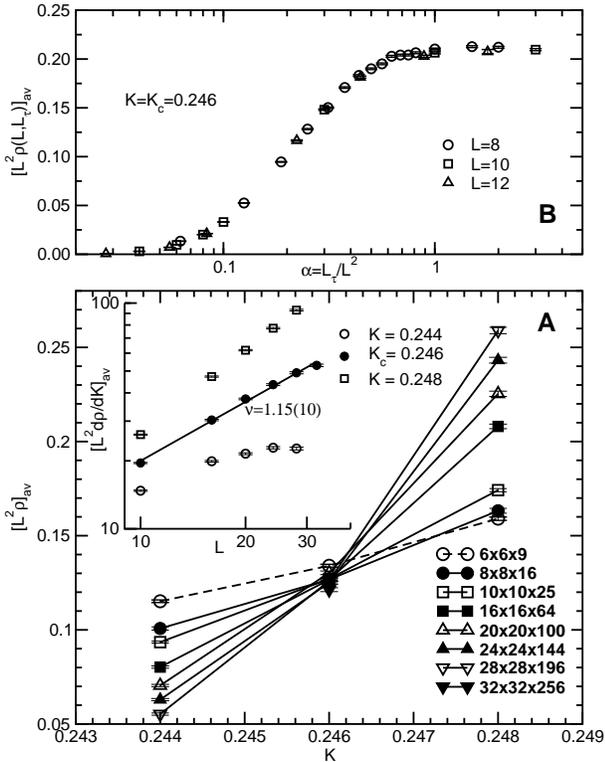}
\caption{({\bf A}) $[L^2 \rho]_{\rm av}$ versus $K$ for different lattice sizes, for $\mu=\frac{1}{2}\pm\frac{1}{2}$. All curves cross
at the critical point $K_c=0.246(1)$ with $[L^2 \rho]_{\rm av}|_{K=K_c}=0.12(1)$.
Inset: $[L^2 d\rho/dK]_{\rm av}$ versus $L$
for different $K$. The solid line indicates a power-law fit yielding an exponent
$\nu=1.15(10)$. ({\bf B}) Scaling plot of $L^2\rho(L,L_\tau)$ at $K_c=0.246$.}
\label{fig:DCross}
\end{center}
\end{figure}
We also simulated the model Eq.~(\ref{eq:hV}) with disorder for $\mu=\frac{1}{2}\pm\frac{1}{2}$. 
In this case the transition is between a superfluid and an insulating boseglass phase.
Scaling theory~\cite{fisher89c} predicts a second order transition with dynamical
exponent $z=2$. Hence, we use lattices of size $L\times L\times \alpha L^2$ where $\alpha=L_\tau/L^2$ is the
aspect ratio.
Previous work~\cite{long}, limited to $L\le 10$, have determined $K_c=0.248\pm0.002$.
Estimates for the correlation length exponent~\cite{long,others} yielded
$\nu=0.9\pm0.1$ almost violating the inequality~\cite{Chayes86} $\nu\ge 2/d$.
From the results shown in Fig.~\ref{fig:DCross} ({\bf A}), obtained with the cluster algorithm,
it is clear that $K_c$ in fact is at a slightly lower value $K_c=0.246(1)$, although
the crossing of $L=6,8$ occurs at $K=0.248$. The disorder average, $[\cdot ]_{\rm av}$, has been
performed over 50,000 samples with 10$^5$ MCS per sample. The more precise value for
$K_c$ significantly changes estimates of $\nu$. The inset in Fig.~\ref{fig:DCross} ({\bf A})
shows $[L^2 d\rho/dK]_{\rm av}$ versus $L$, which at $K_c$ yields $\nu=1.15(10)$ now largely
satisfying the inequality $\nu\ge 2/d$. The results in Fig~\ref{fig:DCross} ({\bf A})
are clearly consistent with $z=2$. In Fig.~\ref{fig:DCross} ({\bf B}) we show results
for $L^2\rho(L,L_\tau)$ versus $L_\tau/L^2$ at $K_c$. Standard scaling theory~\cite{Guo} predicts that this
should be a universal function of $\alpha$ {\it if} $z=2$. Our results
nicely confirm this. The values of exponents are in good agreement with the
analytical estimates in ref.~\cite{Herbut}.

In conclusion, we have introduced a worm algorithm for the quantum
rotor model.  For the link-current representation 
of the quantum rotor model the proposed algorithm is exponentially more
efficient than conventional algorithms and performs at par
with the Wolff algorithm~\cite{wolff} for the classical 3D $XY$ model.
Most noteworthy, the algorithm performs exceptionally well on disordered
systems. We have also successfully adapted it to the study of systems with
longer range interactions as well as classical Ising models~\cite{muquarter}.

We thank M. Troyer and H. Rieger for useful discussions and IDRIS (Paris), CALMIP (Toulouse) and SHARCNET (Hamilton) for time
allocations on their computer systems.

\end{document}